\documentclass[%
 aip,
 sd,%
 amsmath,amssymb,
 reprint,%
]{revtex4-1}

\usepackage{graphicx}
\usepackage{dcolumn}
\usepackage{bm}

\usepackage{subcaption}
\newcommand{\lmatcc}{\left[\!\!\begin{array}{cccc}}
\newcommand{\rmat}{\end{array}\!\!\right]}

\begin{document}

\preprint{AIP/123-QED}

\title{High-Q Contacted Ring Microcavities with Scatterer-Avoiding ``Wiggler'' Bloch Wave Supermode Fields}

\author{Yangyang Liu}\email{yangyang.liu@colorado.edu}

\author{Milo\v{s} A. Popovi\'{c}}\email{milos.popovic@colorado.edu}
\affiliation{Nanophotonic Systems Laboratory, Department of Electrical, Computer, and Energy Engineering,\\University of Colorado, Boulder, CO  80309, USA}

\date{\today}

\begin{abstract}
High-Q ring resonators with contacts to the waveguide core provide a versatile platform for various applications in chip-scale optomechanics, thermo- and electro-optics.  We propose and demonstrate a novel approach to implement azimuthally periodic contacted ring resonators based on multi-mode Bloch matching that support contacts on both the inner and outer radius edges with small degradation to the optical Q.  Radiative coupling between degenerate modes of adjacent transverse spatial order leads to imaginary frequency (Q) splitting and a scatterer avoiding high-Q ``wiggler'' supermode field.  We experimentally measure Q's up to $258,000$ in devices fabricated in a silicon device layer on buried oxide undercladding, and up to $139,000$ in devices fully suspended in air using an undercut step.  Wiggler supermodes are true modes of the microphotonic system that offer new degrees of freedom in electrical, thermal and mechanical design.

\end{abstract}

\maketitle

\noindent Microcavities with attachments have important applications in integrated photonics.  Ridge waveguides extend a flange of the core into the cladding, permitting electrical contacts and good optical confinement to enable electro-optics modulators \cite{Lipson:05}.  Suspended microcavities relying on pedestals \cite{Vahala:04} or inner spokes \cite{Sun:11} for mechanical support have been used in optomechanics. Many of these approaches have limitations: ridge waveguides require a partial etch step, not available for example in standard CMOS processes \cite{Orcutt:12,Shainline13wigglerModulator} 
nor suitable for suspended structures.  Pedestal and inner-spoke cavities either rely on disk-like whispering gallery modes and provide limited degrees of freedom for mechanical design, or have external contacts \cite{Shainline:10,Lipson:05} at the expense of substantial degradation in Q factor due to scattering.


Recently an approach has been proposed to design multimode linear waveguides and resonators with periodic attachments that support modes whose fields avoid those attachments so as to maintain low propagation loss \cite{Popovic07xing,Liu14OL}.  
High-Q resonators based on linear, periodically contacted waveguides that support low-loss ``wiggler modes'' were recently demonstrated \cite{Shainline13cleowiggler,Shainline13wigglerModulator}. 
However, these cavities are large and require non-adiabatic tapers for single-mode to wiggler-mode transitions that may limit their loss Q.

\begin{figure}[ht]

\begin{tabular}{p{.06\linewidth}p{.94\linewidth}}
(a)&
\begin{subfigure}[t]{\linewidth}
\raisebox{1em}{\raisebox{-\height}{\includegraphics[width=.7\linewidth]{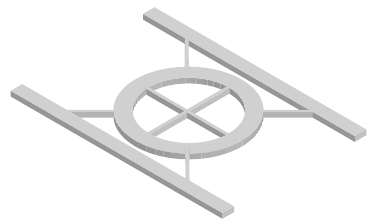}}}
\end{subfigure}\phantomsubcaption\label{sketch}
\end{tabular}

\vskip5pt

\begin{tabular}{p{.06\linewidth}p{.94\linewidth}}
(b)&
\begin{subfigure}[b]{\linewidth}
\raisebox{1em}{\raisebox{-\height}{
\begin{tabular}{ccccccccccc}
\includegraphics[width=.33\linewidth,trim=7cm 2cm 7cm 2cm,clip]{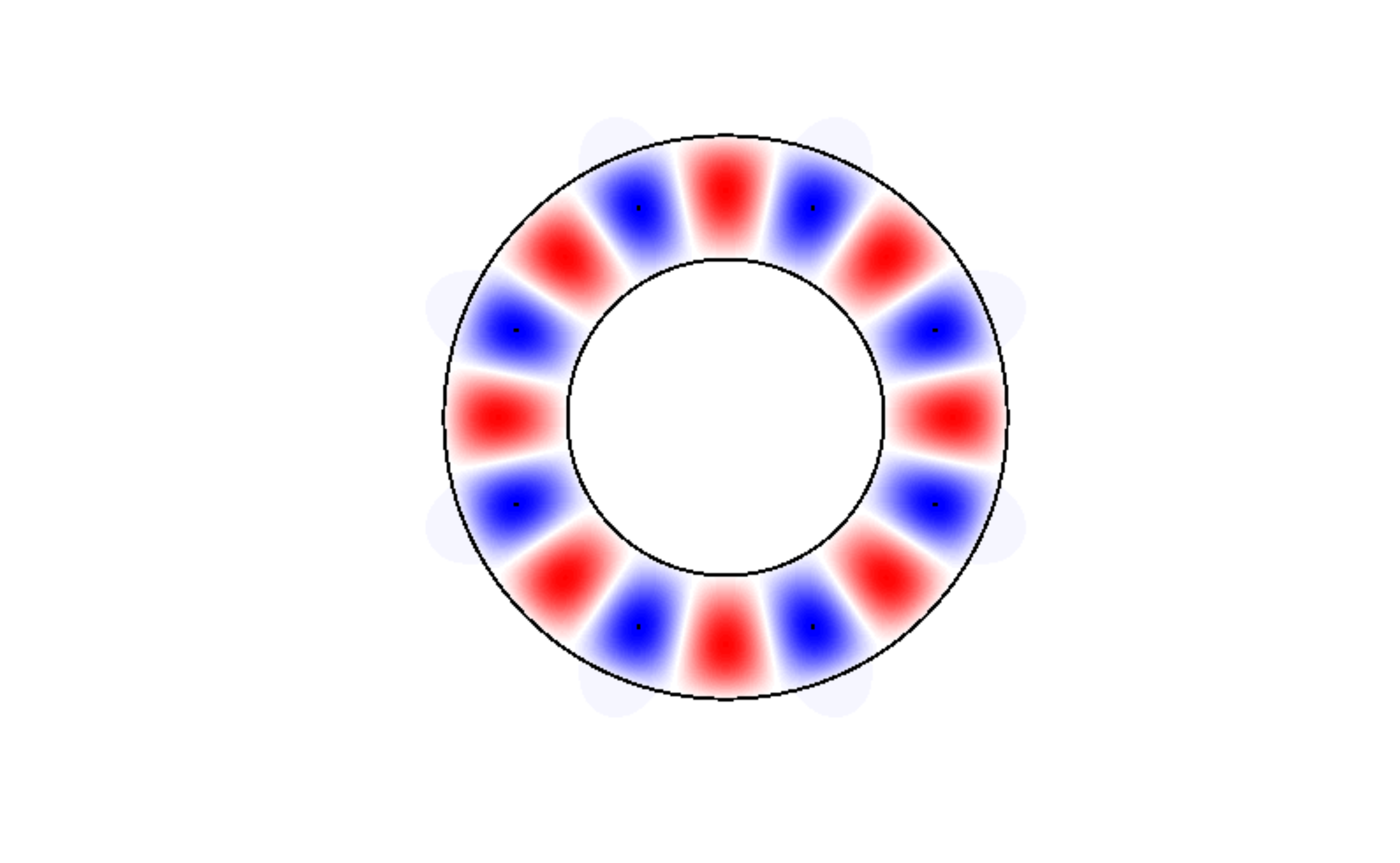}&\raisebox{1.2cm}{\parbox{.5cm}{\centering$+$}}&
\includegraphics[width=.33\linewidth,,trim=7cm 2cm 7cm 2cm,clip]{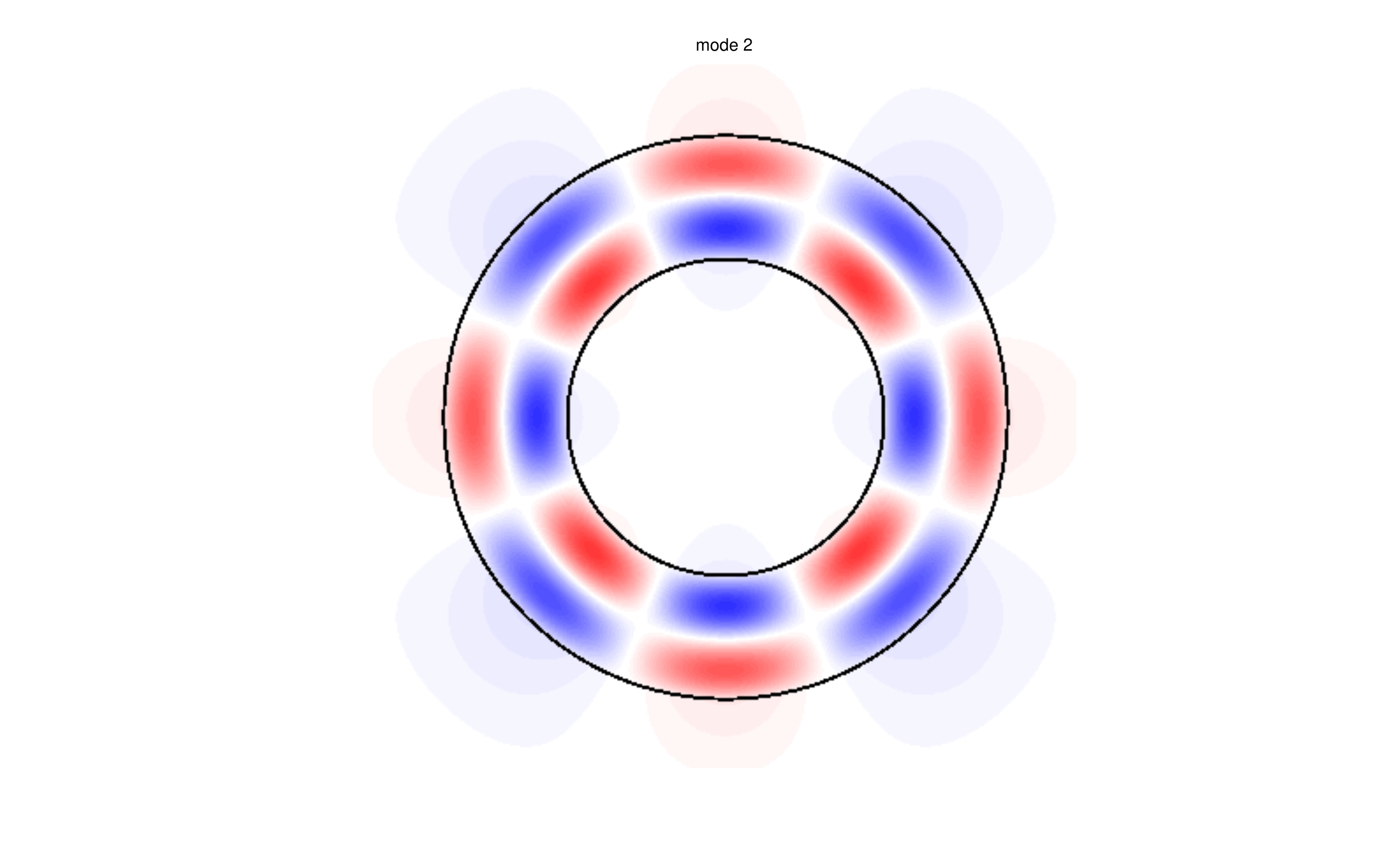}&\raisebox{1.2cm}{\parbox{.5cm}{\centering$\to$}}\\
mode 1 field&&mode 2 field&&\\
\ \\\ \\
\includegraphics[width=.33\linewidth,trim=7cm 2cm 7cm 2cm,clip]{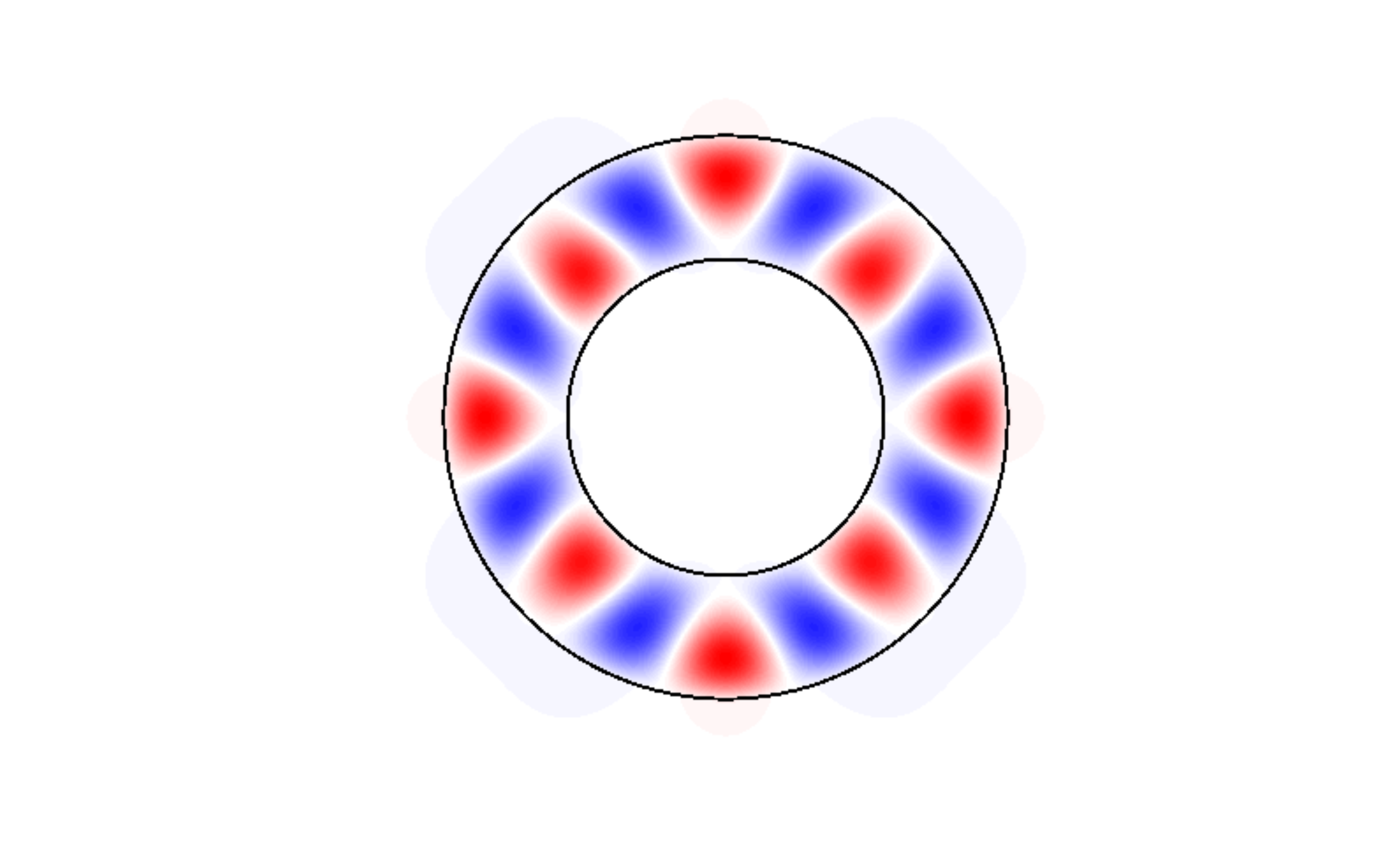}&&
\includegraphics[width=.33\linewidth,trim=7cm 2cm 7cm 2cm,clip]{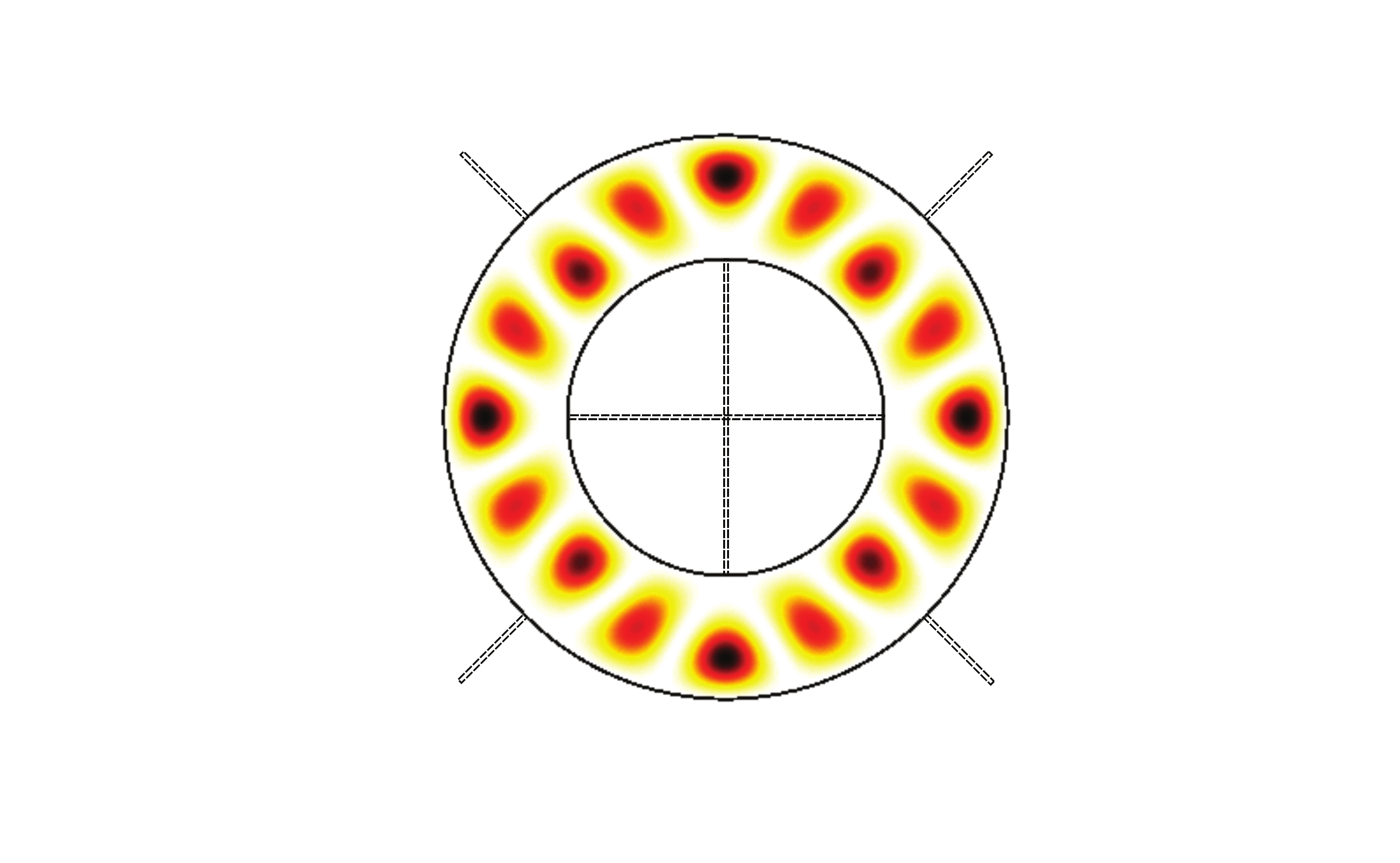}\\
supermode&&supermode\\
$\gamma_1=8$&&$\gamma_2=4$\\
(field)&&(intensity)\\
\end{tabular}}}
\end{subfigure}\phantomsubcaption\label{coupling}
\end{tabular}

\label{fig:concept}
\caption{\protect\subref{sketch} Illustration of proposed microcavity with azimuthally periodic contacts. \protect\subref{coupling} Two degenerate-frequency eigenmodes of an unperturbed ring (1$^\mathrm{st}$-order transverse mode with azimuthal mode number $\gamma_1 = 8$, and 2$^\mathrm{nd}$-order transverse mode with $\gamma_2 = 4$).  A superposition of these modes, which form a low-loss supermode under the scattering perturbation, has a vanishing field intensity at the contact points.}
\end{figure}

In this paper, we demonstrate azimuthally periodic contacted ring microcavities comprising a multimode microring waveguide with periodic attachments to the inner and/or outer walls (Fig.~\ref{sketch}).  These cavities implement a circular symmetry version of the structural Bloch matching and complex Q-splitting concept previously demonstrated in a non-resonant linear waveguide geometry \cite{Liu14OL,Shainline13cleowiggler,Shainline13wigglerModulator}. As in the linear devices, the attachments act as perturbations that lead to a radiative coupling (as opposed to the usual reactive coupling between coupled structures) that splits the first and second fundamental transverse eigenmodes (of different azimuthal order and initially degenerate) of the ring in \emph{imaginary frequency}.  This results in a ``wiggler'' supermode with a field whose spatial distribution avoids the scattering attachments that contact the core, preserving a high Q (Fig.~\ref{coupling}), and another one with high scattering loss at the contacts and thus low Q.
This concept has many applications, allowing electrically and thermally contacted resonators with great freedom in contact geometry and reasonably high optical Q. Further releasing of the devices from the substrate enables thermal isolation and mechanical suspension, which may be useful for thermo-optic effects and thermal isolation design as well as tuning of mechanical properties and optomechanics (light-forces-based devices on chip). We experimentally demonstrate silicon-core ring microcavities with $N=6$ contacts to the core at the outer radius, and show a low-loss resonant supermode with a Q of $258,000$, and an air-suspended ring with $N=4$ contacts and a measured Q of $139,000$.  

\renewcommand{\tabcolsep}{0cm}


To construct a contacted ring resonator that supports a high-Q ``wiggler'' supermode, we start with a single ring that has resonant modes with first- and second-order transverse fields. Interference between these two modes allows the resulting supermode to have a field intensity pattern that wiggles back and forth between the inner and outer walls along the ring (Fig. \ref{coupling}), with a periodicity determined by the difference in azimuthal mode orders.   If an array of contacts is then introduced to the ring, the resulting scattering loss at the contact points can be seen as a perturbation of imaginary permittivity to the original structure, which couples and splits the two modes in imaginary eigenfrequency (or equivalently in loss Q). As in real splitting, the coupling is maximized when the phase matching condition is met, i.e. when the periodicity of the contact array equals the difference in azimuthal mode numbers of the first- and seconds-order transverse modes. 
This loss avoidance behavior of the high-Q ``wiggler'' mode can be understood as the excitation of the right relative amplitudes of the first-order and second-order transverse resonant modes to produce a null in the beat pattern (spatial envelope) of the total electric field at the contact points. 

\begin{figure}[tb]

\begin{tabular}{p{.06\linewidth}p{.94\linewidth}}
(a)&
\begin{subfigure}[t]{\linewidth}
\raisebox{1em}{\raisebox{-\height}{\includegraphics[width=\linewidth]{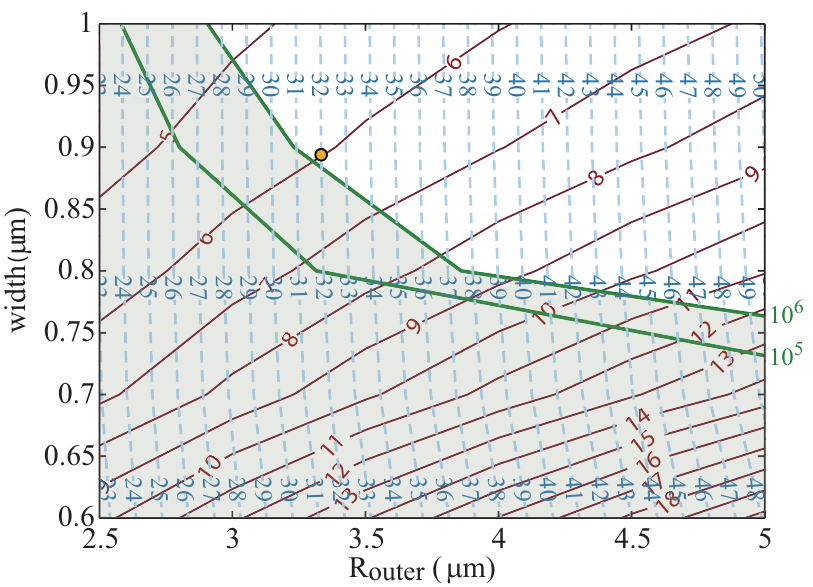}}}
\end{subfigure}\phantomsubcaption\label{map}
\end{tabular}
\vskip5pt

\begin{tabular}{p{.06\linewidth}p{.94\linewidth}}
(b)&
\begin{subfigure}[t]{\linewidth}
\raisebox{1em}{\raisebox{-\height}{\includegraphics[width=\linewidth]{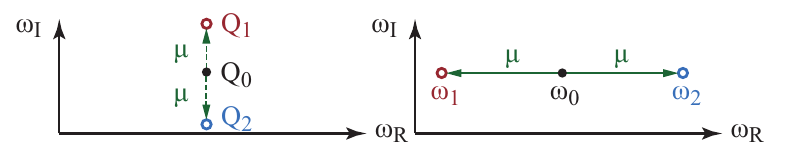}}}
\end{subfigure}\phantomsubcaption\label{splitting}
\end{tabular}
\vskip5pt

\begin{tabular}{p{.06\linewidth}p{.94\linewidth}}
(c)&
\begin{subfigure}[t]{\linewidth}
\raisebox{1em}{\raisebox{-\height}{\includegraphics[width=\linewidth]{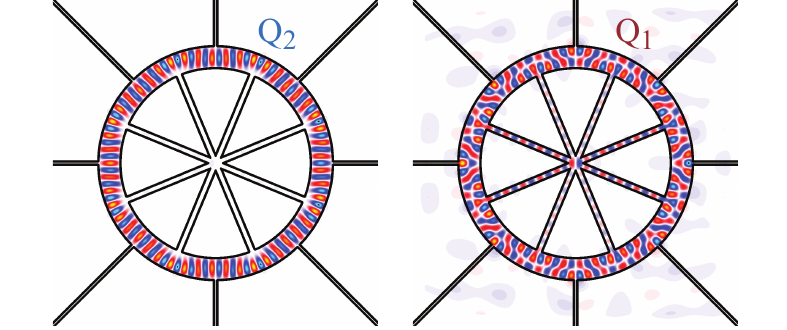}}}
\end{subfigure}\phantomsubcaption\label{modes}
\end{tabular}

\caption{\protect\subref{map} Azimuthal mode number ($\gamma_1$) of 1st order transverse mode of an unperturbed ring at $1550\,$nm (blue, dashed), difference between azimuthal mode numbers of 1st and 2nd order transverse modes ($\gamma_1-\gamma_2$, red), and bending loss Q of the 2$^\mathrm{nd}$ transverse mode (green, thick) versus ring width and outer radius. \protect\subref{splitting} Imaginary (left) and real (right) splitting of initially degenerate modes. \protect\subref{modes} High-Q (left) and low-Q (right) supermode fields resulting from radiative coupling by ring attachments.}
\label{fig:design}
\end{figure}

For experimental demonstration, microcavities were designed in a 220\,nm-thick silicon device layer of typical custom silicon-on-insulator (SOI) wafers for photonics \cite{epixfab}.  
Figure~\ref{map} plots the azimuthal mode numbers of the first-order transverse mode ($\gamma_1$, blue dashed lines), and the difference between azimuthal mode numbers ($\gamma_1-\gamma_2$, red solid lines) of the first and second transverse modes in an unperturbed ring cavity without contacts. At crossing points of the red and blue contours, $\gamma_1$ and $\gamma_2$ are integers and both transverse modes are resonant at the design wavelength of $1550\,$nm.  Bending loss further restricts the dimensions at which high-Q modes can be obtained.  Requiring high-Q initial (uncoupled) modes limits the design space to the upper right part of Fig.~\ref{map}, for example beyond contours of constant bending loss Q of $10^5$ or $10^6$, given as examples in the figure.  The orange marker shows an example design, with $\gamma_1 = 32$, $\gamma_2=26$, and the bend-loss Q of the second mode without attachments just above $10^6$.

An azimuthally periodic array of contacts is then introduced, with periodicity equal to the beat length between the two degenerate modes -- the number of attachment periods is equal to the difference in azimuthal orders of the two modes ($N=\gamma_1-\gamma_2=6$).  These attachments force the resonator into radiative splitting along the imaginary axis on the complex frequency plane (Fig.~\ref{splitting}, left).  This is in contrast to direct reactive coupling of resonators, which leads to real frequency splitting (Fig.~\ref{splitting}, right).  Since quality factor is related to the imaginary part of the complex resonant frequency [$Q \equiv \omega_R/(2 \omega_I)$], the imaginary splitting leads to a high-Q supermode whose field distribution spatially avoids the scattering contacts (illustrated in Fig.~\ref{modes}, left), and a complementary low-Q supermode whose field distribution has strong overlap with the contacts (Fig.~\ref{modes}, right).

\begin{figure}[tb]
\begin{tabular}{p{.06\linewidth}p{.94\linewidth}}
(a)&
\begin{subfigure}[t]{\linewidth}
\raisebox{1em}{\raisebox{-\height}{\includegraphics[width=\linewidth]{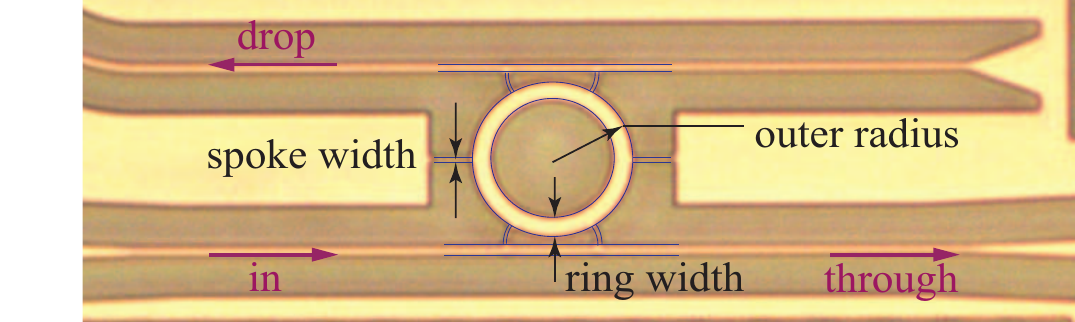}}}
\end{subfigure}\phantomsubcaption\label{unsuspendedOM}\\
\end{tabular}
\vskip5pt
\begin{tabular}{p{.06\linewidth}p{.94\linewidth}}
(b)&
\begin{subfigure}[t]{\linewidth}
\raisebox{1em}{\raisebox{-\height}{\includegraphics[width=\linewidth]{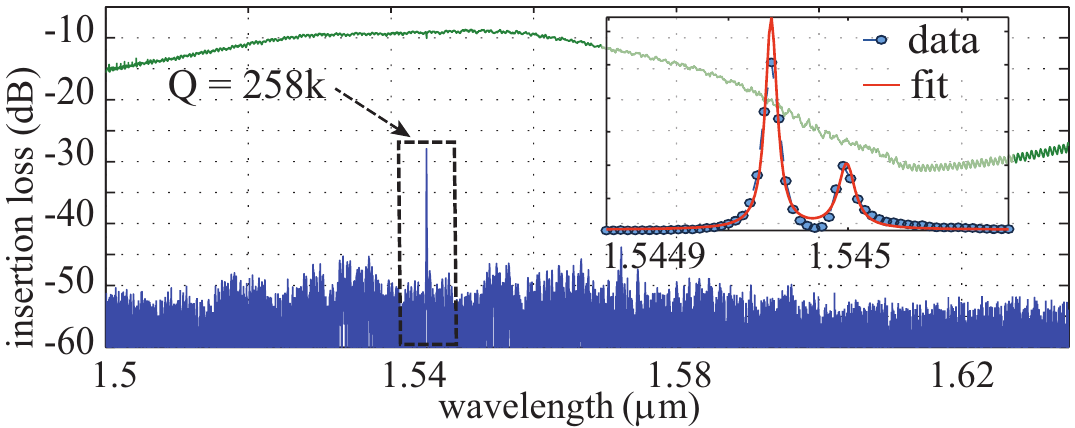}}}
\end{subfigure}\phantomsubcaption\label{hiQ}\\
\end{tabular}
\begin{tabular}{p{.06\linewidth}p{.94\linewidth}}
(c)&
\begin{subfigure}[t]{\linewidth}
\raisebox{1em}{\raisebox{-\height}{\includegraphics[width=\linewidth]{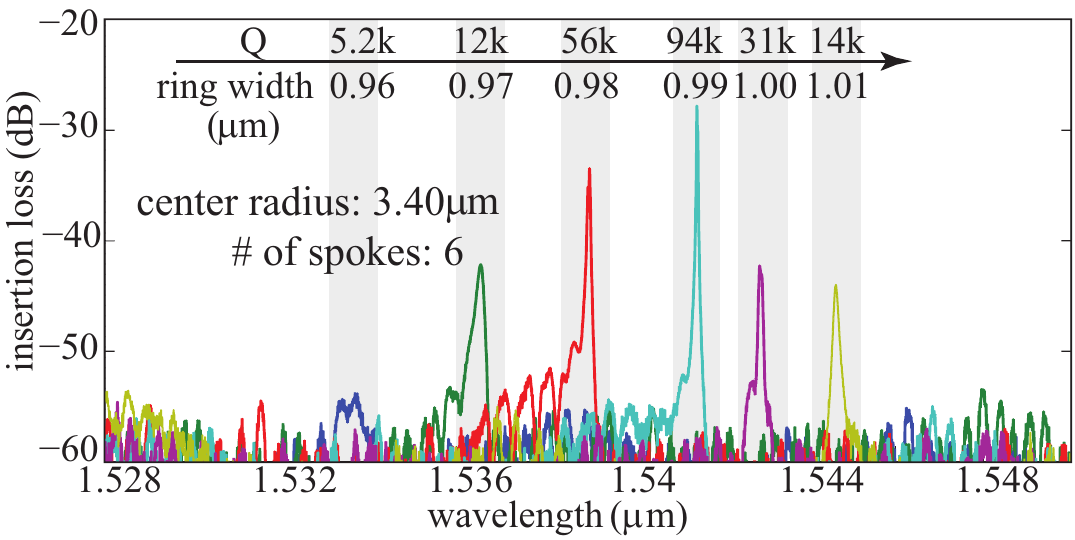}}}
\end{subfigure}\phantomsubcaption\label{spec}
\end{tabular}
\begin{tabular}{p{.06\linewidth}p{.94\linewidth}}
(d)&
\begin{subfigure}[t]{\linewidth}
\raisebox{1em}{\raisebox{-\height}{\includegraphics[width=\linewidth]{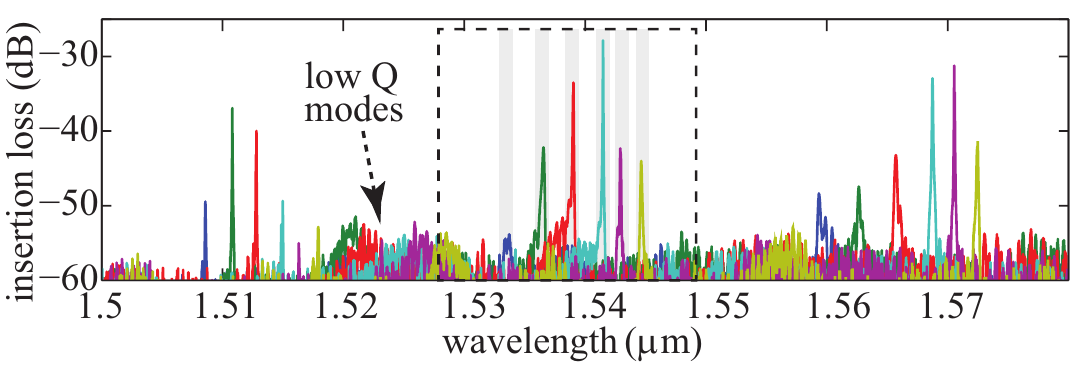}}}
\end{subfigure}\phantomsubcaption\label{spec_all}
\end{tabular}
\caption{\protect\subref{unsuspendedOM} Microscope image of a fabricated ring resonator with 6 attachments. \protect\subref{hiQ} Measured through- and drop-port spectral responses. Inset: resonance with a Q factor of $258,000$. \protect\subref{spec} High-Q resonances of 6 devices with linear width variation show an optimum design with maximum Q factor. \protect\subref{spec_all} Zoomed out view of c showing wider spectra.}
\label{fig:cladded}
\end{figure}

Figure \ref{unsuspendedOM} shows an optical micrograph of a fabricated device with $N=6$ contacts attached to the outer wall of the resonator.  An array of devices was designed around the optimum parameters to account for fabrication variations and to show that a particular combination of radius and width is required for high-Q operation.  The best-performing device was designed with a target outer radius of $3.89\,\mu$m, (multimode) ring cross-section of $990\times220$\,nm, and contact width of 100\,nm.  For this device, Fig.~\ref{hiQ} shows the highest measured Q factor of $258,000$.  The observed resonance doublet is a result of the typical splitting of high-Q traveling-wave modes due to contra-directional coupling (note that here four modes are near-degenerate prior to inclusion of the attachments).  In this case, a lack of azimuthal invariance of the structure contributes enhanced contra-directional coupling.  Note that the drop port is over 15\,dB beneath the through port transmission, which means that the total Q is dominated by the cavity loss Q, not the external coupling \cite{Popovic07phd}.  While the highest observed Q was 258k, many devices with a loss Q over 100k were measured.

Evidence that the thin contacts visible in Fig.~\ref{unsuspendedOM} are not inconsequential, i.e. that our design approach is necessary to obtain high Q, is provided in Fig.~\ref{spec}.  The figure shows drop-port spectra of representative devices with a fixed waveguide-cavity coupling gap ($250$\,nm) and radius ($R_\mathrm{center} = 3.4\,\mu$m), while the width is linearly varied (see Fig.~\ref{spec}).  The resonances red-shift with increased width due to increase in effective index.  However, the transmission and the Q increase (linewidth decreases) as one approaches the center device from either side.  These both confirm that the highest loss Q is in the central device, as the waveguide coupling is broadband and does not contribute to variation in insertion loss from one resonance to the next.  In addition, these spectra show very broad, low transmission resonances interspersed between the high Q ones (e.g. see broad resonances in Fig.~\ref{spec_all}).  These resonances represent the complementary, low-Q resonant supermodes that result from the imaginary frequency splitting.

%
%
%

\begin{figure}[tb]
\includegraphics[width=.85\linewidth]{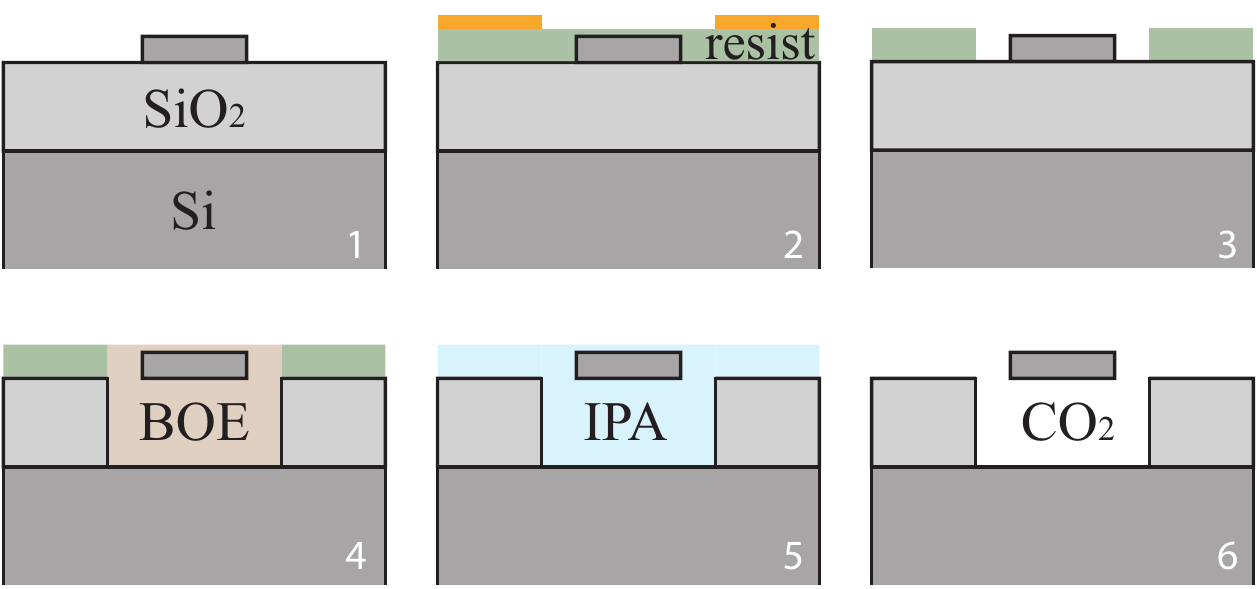}
\caption{Process flow for suspending the resonators. A positive resist is spin on the substrate and patterned with a chromium mask to serve as etch windows; oxide under the resonator is then selectively etched with a buffered HF solution (BOE); liquid under device layer is replaced with IPA and eliminated in a CO$_2$ critical point dryer.}
\label{fig:process}
\end{figure}

A different device designed for air cladding on both sides of the silicon layer was fabricated on the same chip and further released in post processing (Fig. \ref{fig:process}).  
A chromium mask is created with etch windows in dark field. A $900\,$nm layer of positive resist (NR9-1000P) is spun onto the chip at $4000\,$rpm for $40\,$s. The coated chip is then aligned to the mask and exposed for $15\,$s. After development, the chip is submerged in buffered HF (BOE) for $20\,$min to selectively remove the oxide ($2\,\mu$m thick) under the resonator region. To avoid sticking of the device to the bottom silicon due to surface tension in normal evaporative drying, after rinsing in DI water, IPA is used to replace the liquid covering the chip, and is eventually removed in a CO$_2$ critical point dryer.

\begin{figure}[t!]

\begin{tabular}{p{.06\linewidth}p{.94\linewidth}}
(a)&
\begin{subfigure}[t]{\linewidth}
\raisebox{1em}{\raisebox{-\height}{\includegraphics[width=.6125\linewidth]{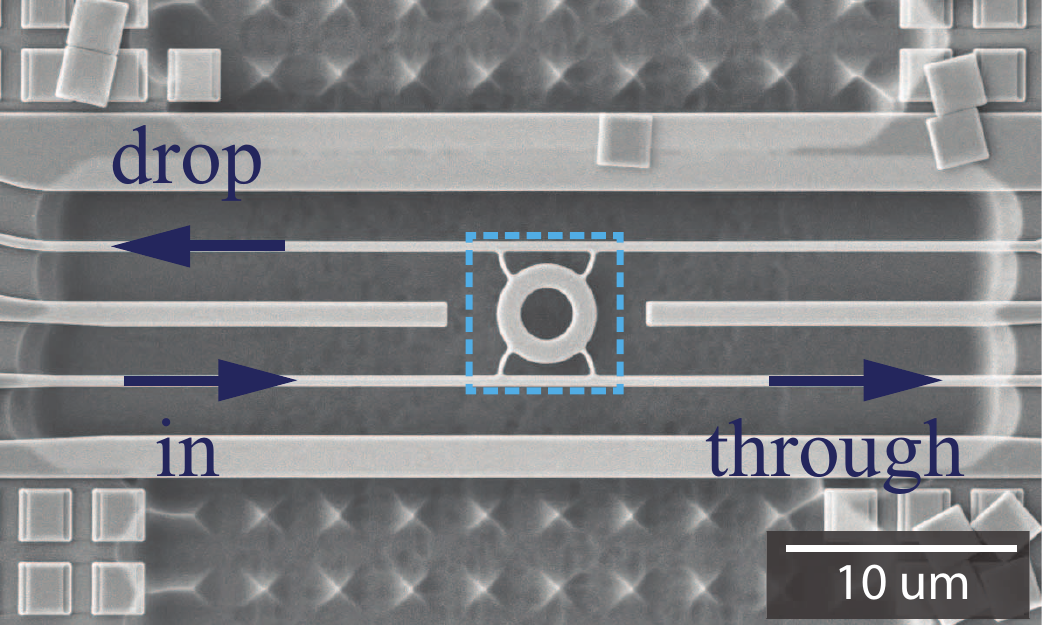}
\hskip1pt
\includegraphics[width=.3675\linewidth]{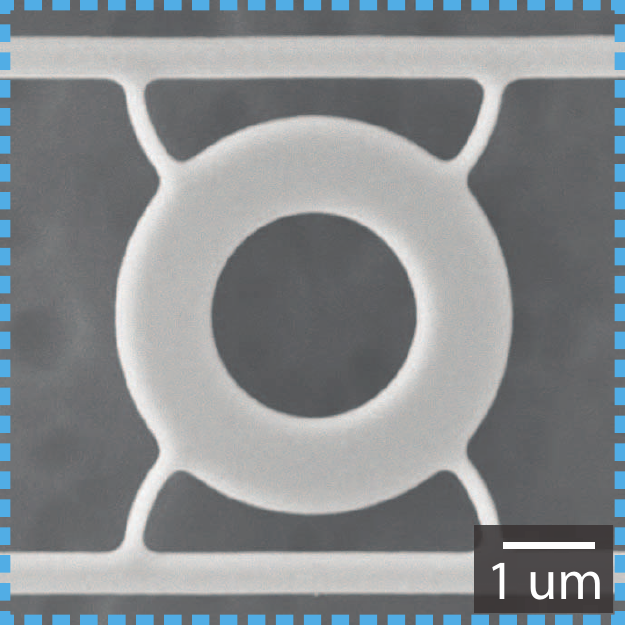}}}
\end{subfigure}\phantomsubcaption\label{undercutSEM}
\end{tabular}

\vskip5pt

\begin{tabular}{p{.06\linewidth}p{.94\linewidth}}
(c)&
\begin{subfigure}[t]{\linewidth}
\raisebox{1em}{\raisebox{-\height}{\includegraphics[width=\linewidth]{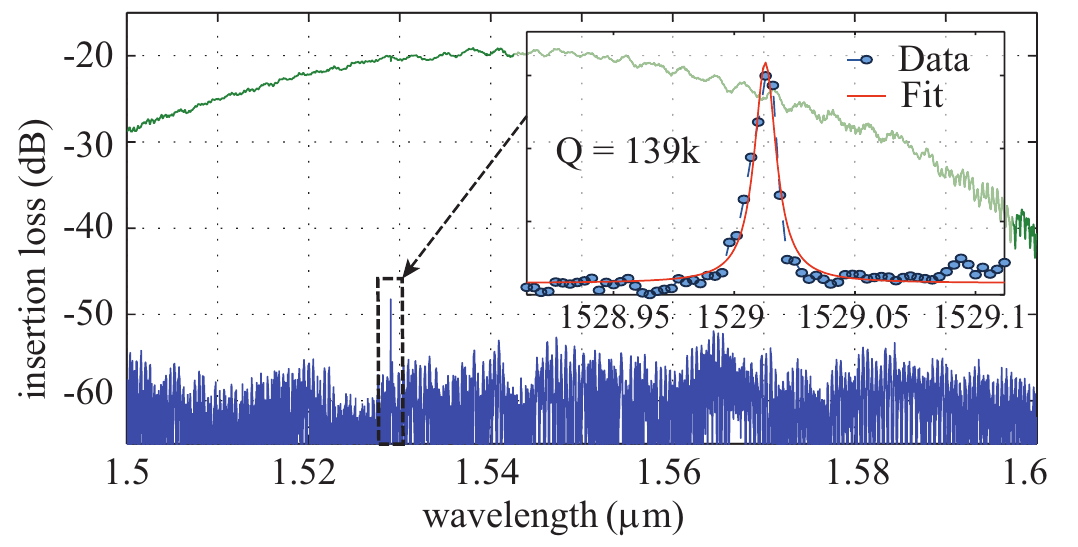}}}
\end{subfigure}\phantomsubcaption\label{undercut_hiQ}
\end{tabular}
\caption{\protect\subref{undercutSEM} SEM images of an air suspended  ring resonator with 4 attachments attached to coupling waveguides (left) and a zoomed in image of the resonator region (right). \protect\subref{undercut_hiQ}~Measured through- and drop-port spectral responses. Inset: resonance with a Q factor of $139,000$.}
\label{fig:undercut}
\end{figure}

Figure \ref{undercutSEM} shows an SEM image of a released structure and a zoomed in picture of the resonator. The device is a ring resonator with 4 contacts connected to an input and an output coupling waveguide, similar to the 6-contact device shown in Fig. \ref{unsuspendedOM}. 
The small squares visible on the four corners are density fill pattern required in the SOI fabrication process for process uniformity \cite{epixfab}.  BOE etching was timed to remove the entire thickness ($2\,\mu$m) of the silica under the device layer, and some fill shapes within the etch window were removed from the chip during the etching and drying, leaving pyramid shaped residuals at their original sites.  A higher magnification SEM image on the right confirms that the air-suspended resonator is supported by its $4$ contacts connecting it to the partially released waveguides.  Direct mechanical connection to a waveguide can be a useful feature for suspended photonic structures, as built in stresses in the device layer can produce out of plane misalignment of adjacent waveguides and other structures without proper stress relief within the design\cite{LoncarBuckling}.  In this paper, we did not design scattering avoiding structures into the waveguide as well, only the resonator, since the waveguide sees only a single pass loss.  However, straight wiggler mode taper arms could be incorporated in the bus waveguide, similar to those in racetrack resonators\cite{Shainline13cleowiggler,Shainline13wigglerModulator}, to minimize overall transmission loss of attached wiggler-mode cavities.

Figure \ref{spec} 
shows measured through- and drop-port spectral responses from a suspended ``wiggler'' resonator with a quality factor of $139,000$, indicating low loss. In comparison to the devices in Fig.~3 that are not suspended,
these structures also show significantly stronger thermal nonlinearity under higher input power.

The proposed wiggler mode concept, incorporating imaginary coupling (frequency splitting), is demonstrated in a circular symmetry geometry with azimuthal periodicity.  The demonstrated resonators may enable important new degrees of freedom in design of photonic and nanomechanical structures by allowing manipulation of electrical function, thermal properties (mass, impedance), nanomechanical design and phonon modes while preserving a high optical Q.  The design may be enabling for applications in light-force actuated photonics on chip, e.g. for state trapping and self-adaptive photonics\cite{Rakich07trapping}, very efficient thermal tuning by allowing mechanical suspension with potentially higher thermal impedance than e.g. microdisks on a pedestal or even wheel resonators, as well as in other applications such as enabling the tighter mode confinement of an air-silicon interface within a microcavity.

This work was supported by NSF grant ECCS-1128709.  Fabrication was done at IMEC, Belgium, through the ePIXfab Multi-Project Wafer silicon photonics shuttle runs (IMEC10).

%

\end{document}